# Primordial Porous Structure of Chondrite Parent Bodies Due To Self-Gravity

Tomomi Omura[1] and Akiko M. Nakamura[2]

[1] Institute of Education Center of Advanced Education, Osaka Sangyo University, 3-1-1 Nakagaito, Daito-shi, Osaka, 574-8530, Japan; tomura@edo.osaka-sandai.ac.jp
[2] Department of Planetology, Kobe University, 1-1 Rokkodai-cho, Nada-ku, Kobe, Hyogo, 657-8501, Japan; amnakamu@kobe-u.ac.jp

## Abstract

The porosity of an asteroid is important when studying the evolution of our solar system through small bodies and for planning mitigation strategies to avoid disasters due to asteroid impacts. Our knowledge of asteroid porosity largely relies on meteorites sampled on Earth. However, chondrites sampled on Earth are suggested to be sorted by strength. In this study, we obtained an estimate of the most porous structure of primordial "granular" chondrite parent bodies based on measurements of the compaction behavior of chondrite component analogs. We measured compaction curves of dust and dust–bead mixture samples. The dust sample consisted of various spherical and irregular particles with diameters on the order of $10^0$–$10^1$ μm. The mixture sample consisted of dust and beads with different dust volume fractions (~0.2–1). We used 1.5 and 4.8 μm particles as dust as a first step, although the typical size of materials in matrix may be much smaller. We approximated the compaction curve of each sample with a power-law form and calculated the porosity structure of the primordial chondrite parent bodies using the experimental results. Our results show that the primordial parent bodies are likely to have higher porosity than the chondrites. Moreover, the relatively higher volume fraction of the matrix may be one of the reasons why most meteorites with high porosity are carbonaceous chondrites.

## 1. Introduction

The physical properties of an asteroid, such as elasticity, strength, and thermal conductivity, reflect and affect its evolution and determine the efficiency of strategies that can be employed to avoid a collision between the asteroid and the Earth (Ahrens & Harris 1992; Stickle et al. 2020). Such properties are dependent on the porosity of the asteroid.

The bulk porosity of asteroids ranges from ~0 to ~0.7 and can be estimated from the bulk density and the grain density of their corresponding meteorites (Baer et al. 2011). The porosity of asteroids primarily originates from two types of internal structures: the primordial microstructure between particles provided by the protoplanetary disk (microporosity) and the macrostructure between reaccumulated rocks, as shown in rubble-pile bodies (macroporosity), which are presumed to be clues to the collisional evolution of asteroids (Britt et al. 2002). Macroporosity is estimated based on assumed microporosity. The microstructure can be preserved in carbonaceous and ordinary chondrites, which are the types of meteorites delivered from C- and S-type asteroids, respectively.

Chondrites consist of submicrometer- to micrometer-sized matrix particles and other coarse materials, such as chondrules; the size and volume fraction of the constituent materials vary depending on the chondrite groups (Weisberg et al. 2006). Additionally, the porosity of chondrites is different for each chondrite group. Porosities primarily range from ~0.1 to 0.35 (Consolmagno et al. 2008), and carbonaceous chondrites tend to have higher porosities than ordinary chondrites (Consolmagno et al. 2008). Some meteorites, including the Tagish Lake meteorite, have even higher porosity, ~0.25–0.5 (Ralchenko et al. 2014). Furthermore, the measured thermal inertia of asteroid Ryugu and a typical boulder on the asteroid indicates that the asteroid consists of more highly porous material (porosity of ~0.3–0.5) than typical chondrites (Grott et al. 2019; Okada et al. 2020). These observations suggest the biasing of chondrites based on their strength (Consolmagno et al. 2008; Scheeres et al. 2015).

The porosity of asteroidal materials is the result of the internal stress state of the body that is due to self-gravity, spin, tidal force, compression due to impact (Hirata et al. 2009; Beitz et al. 2013), and thermal and aqueous alteration. An estimate of the most porous structures of the chondrite parent bodies can be obtained by calculation of the compaction of the primordial particles that is due to the self-gravity of the bodies using the compaction behavior of the materials approximated by a power-law form (Omura & Nakamura 2018). The compaction behavior of the constituent materials of a chondrite parent body, which affects the structures of the body, vary depending on the chondrite group, that is, the volume fractions of the matrix (fine dust) and the chondrules (millimeter to submillimeter spheres).

In this study, we conducted compaction experiments using chondrite component analogs (dust–bead mixture) to investigate the effect of the particle size of dust and the mixing ratio of components on their compaction behavior. We then obtained an estimate for the most porous structure of chondrite parent bodies based on their compaction curves.

## 2. Laboratory Measurements

### *2.1. Sample*

Various spherical and irregular dust particles with diameters on the order of $10^0$–$10^1$ μm were used in this study. The major components of the dust particles were $SiO_2$, $Al_2O_3$, or serpentinite; characteristics of the sample powders are shown in Table 1. The size distribution of the dust particles, shown in Figure 1, was measured using a laser diffractometer (Shimazu SALD-3000S) installed at Kobe University. Here, the particle size in which the cumulative volume fraction becomes $x$% was referred to as $d_x$. The median diameter ($d_{50}$) and $d_{85}/d_{15}$ ratio were used as the indicators of the particle diameter and size distribution of each dust sample, respectively. The mixture samples consisted of dust and beads with different dust volume fractions (~0.2–1). Silica powder (irregular shape) with a median diameter of 1.5 μm and fly ash (spherical shape) with a median diameter of 4.8 μm were selected as dust materials, based on the particle sizes of the Allende matrix, which vary depending on the measurement site and range from ~1 nm to ~10 μm ( Toriumi 1989; Forman et al. 2019). The average particle diameter measured at four sites using scanning electron microscopy was ~1 μm. Glass beads with a diameter of 1 mm were used as chondrule analogs (this size is similar to the average chondrule diameter of CV-group chondrites; Weisberg et al. 2006).

However, the most primitive dust materials in the extraterrestrial sample collection are the chondritic porous interplanetary dust particles (CP IDPs). CP IDPs are considered to originate from comets based on their porous and fragile structures and mineralogical and chemical characteristics and do not show the evidence of significant alteration on their parent body (Bradley 2014). Therefore, CP IDPs may be a sample of the primitive matrix before alteration. The vast majority of the individual grains in the CP IDPs had a size of ~50–200 nm (Wozniakiewicz et al. 2013). In situ measurement of particles from comet 67P revealed that they have a composition and morphology similar to CP IDPs (Hilchenbach et al. 2016) and consist of submicron-sized individual grains with structures that indicate hierarchical aggregation (i.e., small aggregates construct larger aggregates; Bentley et al. 2016). The particle size is significantly smaller than that of our samples and should affect the compaction behavior of the chondrite parent body. The effect of particle characteristics on the compaction behavior is discussed in Section 3.

Sample containers were stainless cylinders with inner diameters and depths of ~60 and ~30 mm, respectively. Dust particles were sieved into the sample container using a 500 μm mesh screen. The mixtures were poured into the sample container using a spoon. The part of the bed that exceeded the height of the container was leveled off with a spatula. A small sample container with inner diameter and depth of ~20 and ~17 mm, respectively, was used for some dust samples because of the longer sample preparation time for highly cohesive powders or the limited amount of sample powders. Powders were kept in the laboratory and were not dried prior to conducting the experiment, with some exceptions. The information for each sample is shown in Tables 2 and 3.

**Table 1 Characteristics of sample particles.**

| Name | Grain Density ($gcm^{-3}$) | Diameter (μm) | $d_{85}/d_{15}$ | Shape | Material |
|---|---|---|---|---|---|
| Silica beads (1.7 μm) | 2.2 | 1.7[a] | 2.8 | Spherical | $SiO_2$ |
| Fly ash (4.8 μm) | 2.0 | 4.8[a] | 6.3 | | |
| Glass beads (18 μm) | 2.5 | 18[a] | 1.4 | | |
| Glass beads (1 mm) | 2.5 | 1000[b] | - | | |
| Silica powder (1.5 μm) | 2.2 | 1.5[c] | 20 | Irregular | |
| Silica sand (13 μm) | 2.645 | 13[a] | 11 | | |
| Silica sand (19 μm) | 2.645 | 19[a] | 22 | | |
| Silica sand (73 μm) | 2.645 | 73[a] | 2.0 | | |
| Serpentinite (10 μm) | 2.55 | 10[a] | 23 | | $(Mg,Fe)_3Si_2O_5(OH)_4$ |
| Alumina (1.0 μm) | 3.85 | 1.0[a] | 4.9 | Irregular | $Al_2O_3$ |
| Alumina (1.7 μm) | 3.94 | 1.7[a] | 14 | | |
| Alumina (1.8 μm) | 3.85 | 1.8[a] | 8.3 | | |
| Alumina (4.5 μm) | 4.0 | 4.5[a] | 11 | | |
| Alumina (6.5 μm) | 3.90 | 6.5[a] | 2.3 | | |
| Alumina (15 μm) | 3.90 | 15[a] | 1.6 | | |
| Alumina (59 μm) | 3.90 | 59[a] | 1.6 | | |
| Alumina (77 μm) | 4.0 | 77[a] | 1.6 | | |

[a] Determined from the measured particle size distribution.

[b] Manufacturer's information, UNITIKA.

[c] Manufacturer's information, MARUTO

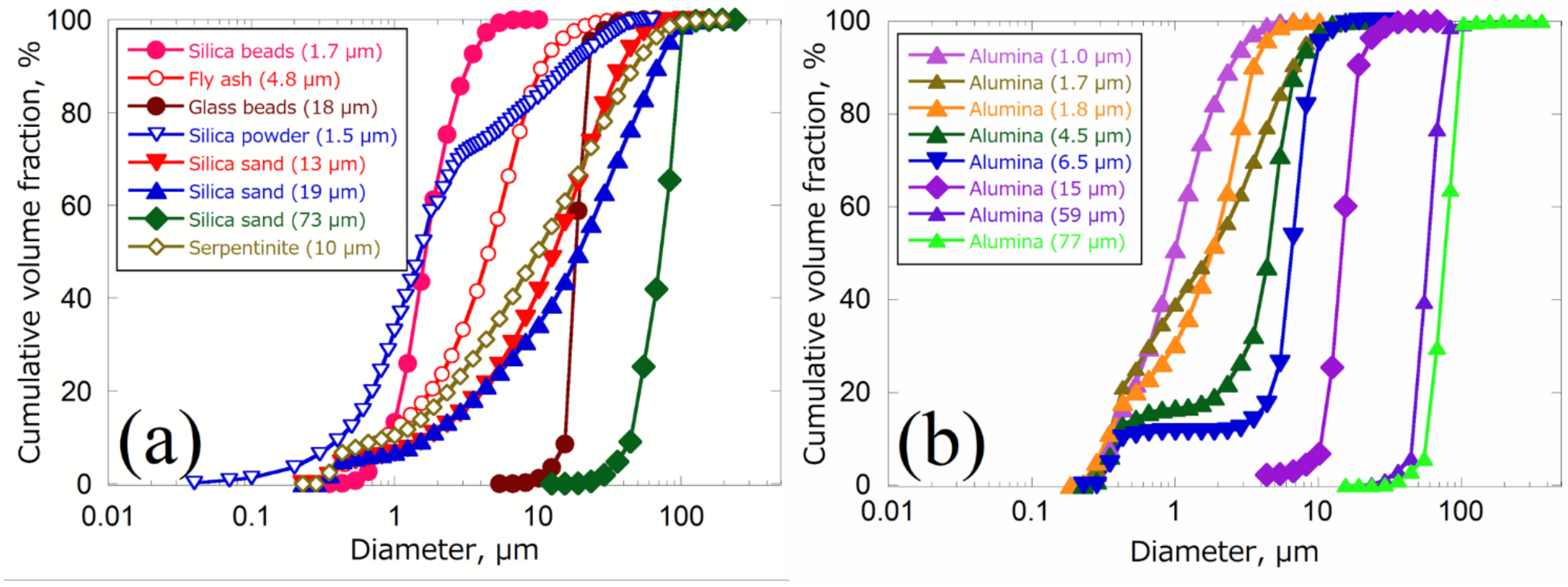

**Figure 1.** Particle size distribution of (a) silica and serpentinite particles and (b) alumina particles. The dust materials used in the mixture samples are shown using blue open triangles and red circles.

*2.2. Experimental Setup*

The results of centrifugal compression (body-force compression) and piston compression were consistent when the density gradient inside the sample was considered (Suzuki et al. 2004). Therefore, compaction experiments were conducted using a piston fixed to a compressive testing machine installed at Kobe University (Shimazu EZ-Graph). The diameter of the piston was slightly smaller than the inner diameter of the sample container, to avoid the effect of friction between the inner wall of the container and the piston. The gap between the piston and inner wall of the container was 0.3–1 and 0.8–3.4 mm for the dust and mixture samples, respectively. The stroke of the piston and the forces required to maintain the constant loading rate were recorded. The pressure applied to the sample was calculated by dividing the compressive force applied to the sample by the cross-sectional area of the piston. The maximum applied pressure was approximately $5 \times 10^6$ Pa, and the loading rate was 10 $\mu m s^{-1}$. Additionally, we conducted an experiment with a loading rate of 1 $\mu m s^{-1}$ for a sample. The resulting compaction curve was almost the same as that obtained at a loading rate of 10 $\mu m s^{-1}$.

*2.3. Calculation of Filling Factor and Porosity*

The relationship between the porosity, $\varepsilon$, and the filling factor, $f$, is

$$\varepsilon = 1 - f. \tag{1}$$

The filling factor of the dust sample was calculated as follows:

$$f = M/\rho_g V, \tag{2}$$

where $M$ is the sample mass, $\rho_g$ is the grain density of the sample particle, and $V$ is the sample volume calculated from the cross-sectional area of the sample container and the sample height estimated from the stroke of the piston.

The filling factor of the mixture sample was calculated from the sample volume and the particle volume (dust + beads), $V_p$, placed under the piston. The particle volume is the sum of the volume of the dust particles and the volume of beads estimated from the total mass of the mixture sample, $M$, and the mass fraction of dust, $D$,

$$V_p = V_D + V_B = \frac{DM}{\rho_D} + \frac{(D-1)M}{\rho_B}, \tag{3}$$

where $V_{D \text{ or } B}$ and $\rho_{D \text{ or } B}$ are the volume and grain density of dust particles or beads, respectively. The apparent grain density, $\rho_a$, is calculated as follows:

$$\rho_a = \frac{M}{V_p}. \tag{4}$$

In some experiments, particularly in the case of the mixture samples, a nonnegligible amount of sample particles remained between the piston and the inner wall of the container, because the piston used for the mixture samples was thinner than that used for dust-only samples. The remaining particle amounts varied according to the experimental conditions (the inner diameter of the sample container, the diameter of the piston, and the sample powder). To correct the volume of particles, we assumed that the particle layer between the piston and the wall was not compacted and remained in its initial position in all cases. However, this assumption probably resulted in a lower filling-factor estimate. Therefore, based on this assumption, the volume of the corrected sample particles placed under the piston $V_p'$ was calculated as follows:

$$V_p' = V_p - f_0(S_c - S_p)(h_c - h), \tag{5}$$

$$f_0 = \frac{M/(S_c h_c)}{\rho_a}, \tag{6}$$

where $f_0$ is the initial filling factor of the sample, $S_{c\text{ or }p}$ is the cross-sectional area of the sample container or piston, $h_c$ is the depth of the sample container, and $h$ is the sample height. Using this particle volume, the corrected sample filling factor was determined as

$$f = \frac{V_p'}{V}. \tag{7}$$

In the case of the mixture samples, the corrected filling factor was used.

*2.4 Results*

The relationship between the pressure applied to a dust–bead mixture sample, which is used as the chondrite component analog, and the porosity of the sample is shown in Figure 2. This relationship shows the data at a pressure lower than $4 \times 10^6$ Pa but higher than the "yield strength" (Omura & Nakamura 2017) of the sample; the initial fluffy structure of the granular sample is retained below this pressure, even when the applied pressure increases because the initial structure is already determined by the static pressure of the gravity of the Earth. Basically, the sample with a lower dust volume fraction has a lower porosity, because there are no pores inside glass beads. In the case of the sample with a dust volume fraction of ~0.2, the compaction behavior is dominated by the configuration of the glass beads, and the porosity of the sample is almost unchanged under a pressure between $10^5$ and $3 \times 10^6$ Pa. Additionally, in the case where two samples have similar dust volume fractions, the sample consisting of smaller dust particles (1.5 μm) has a higher porosity and is easier to compact.

We introduced a modified two-layer model to measure the effect of the dust volume fraction on the compaction curve. This model is based on a two-layer model (Yasui & Arakawa 2009). We assumed that the beads contained in the sample do not affect the compaction behavior of the dust. Instead, they are regarded as a solid layer that has the same volume as the sum of the volume of each grain and are not compressive. The porosity of the mixture under pressure, $P$, $\varepsilon_m(P)$, was obtained by the filling factor of the mixture under pressure, $P$, $f_m(P)$, as follows:

$$\varepsilon_m(P) = 1 - f_m(P) = 1 - \left\{\frac{V_D + V_B}{(V_D/f_D(P)) + (V_B/f_B(P))}\right\} = 1 - \left\{\frac{V_D + V_B}{(V_D/f_D(P)) + V_B}\right\}, \tag{8}$$

where $f_{D\text{ or }B}(P)$ is the filling factor of the dust or beads under pressure $P$; $f_D(P)$ was given by the compaction curve of the sample with a dust volume fraction of 1. We assumed that $f_B(P)$ did not change with the applied pressure and was always unity. This model ignores the proportion of sample particles remaining between the piston and the inner wall of the container.

The compaction curves calculated according to the modified two-layer model are shown in Figure 2 and were used to evaluate the effect of the compaction behavior of the dust on that of the mixture sample. The curves are consistent with the experimental results for samples with dust volume fractions larger than ~0.5, suggesting that the compaction behavior of chondrite parent bodies is dominated by that of the matrix when the matrix volume fraction is larger than ~0.5. The compaction behavior of the powder layer is affected by the surface conditions of the individual particles (Castellanos et al. 2005). Individual grains in CP IDPs are coated with ~100 nm thick organic matter before their aggregation (Flynn et al. 2013). This organic layer may work as a glue or grease, and the compaction behavior of the chondrite parent body will be affected. The matrix may have contained ice particles; the water content of the matrix is 3-18 wt% for an extremely primitive carbonaceous chondrite (Matsumoto et al. 2019). The amount of ice and the size of the particles within the matrix are generally poorly constrained. However, the compaction behavior would have been affected by the presence of abundant ice particles or particle sizes small enough to cover the matrix particles.

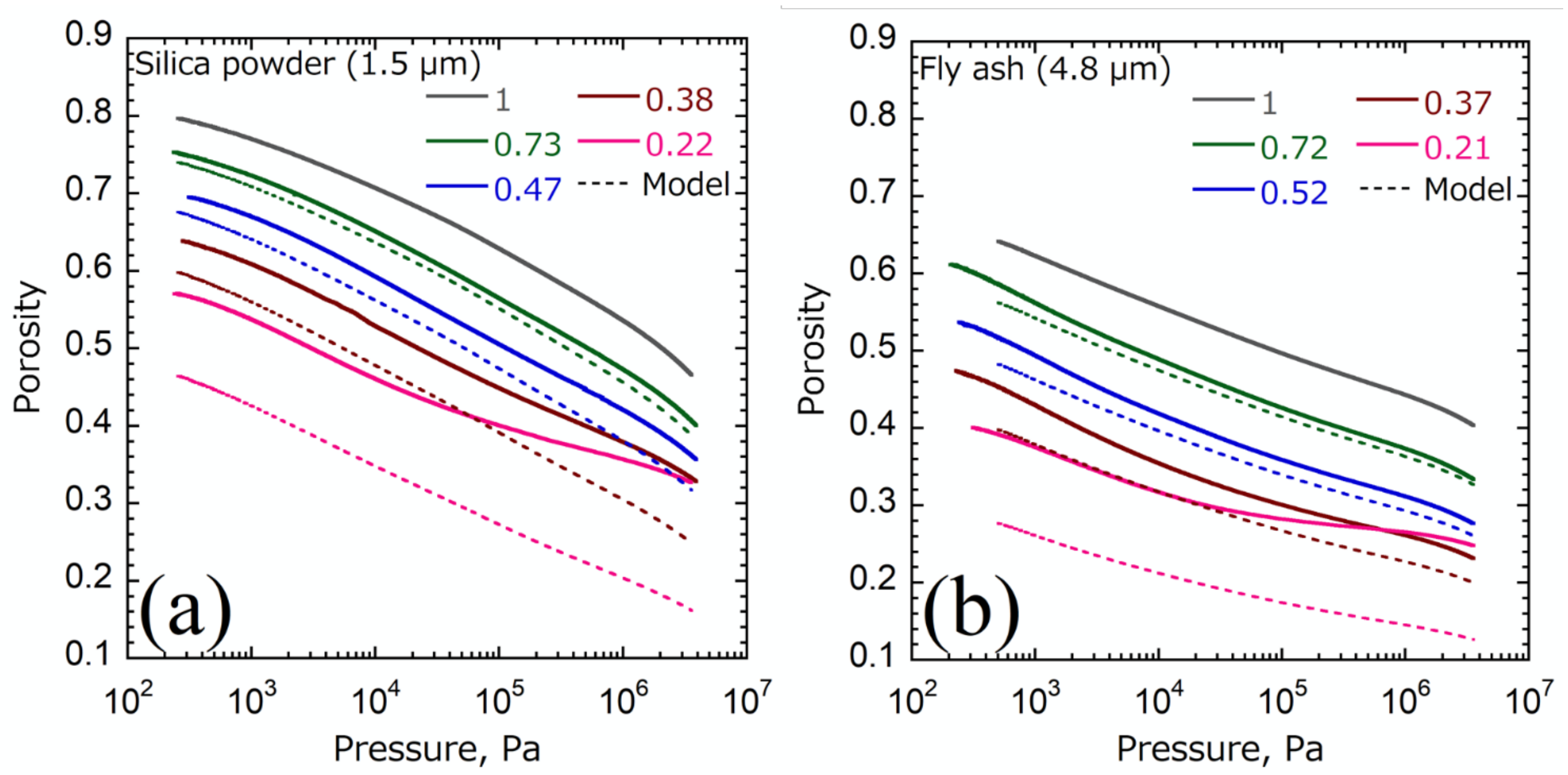

**Figure 2**. Compaction curves of mixture samples obtained by measurements and calculated using the modified two-layer model. The results of samples consisting of (a) silica powder (1.5 μm) and (b) fly ash (4.8 μm) are shown. The experimental results are shown by solid curves. The color of each curve corresponds to the dust volume fraction, which is indicated by numbers. A typical compaction curve of three measurements is shown, except for in the cases of samples with a dust volume fraction of 1. The error bar corresponding to the uncertainty of the sample height is smaller than the thickness of the curves. Dotted curves show the calculation results obtained using the modified two-layer model.

**Table 2 Information and Results of Dust Samples**

| | | | | Sample Container | | Piston | | | Fitting Range | | |
|---|---|---|---|---|---|---|---|---|---|---|---|
| No. | Name | Dried Particles Yes/No | Preparation Method | Inner Diameter (mm) | Depth (mm) | Diameter (mm) | *K'* | *n* | Min (Pa) | Max (Pa) | Reference |
| D150310-6 | Silica sand (13 μm) | No | Sieving | 58 | 33 | 57.7 | $(4.95\pm0.49) \times 10^{12}$ | $(4.437\pm0.025) \times 10^{-2}$ | $1.5 \times 10^4$ | $3.9 \times 10^5$ | 1 |
| D150310-1 | Silica sand (19 μm) | No | Sieving | 58 | 33 | 57.7 | $(6.32\pm0.45) \times 10^{11}$ | $(4.244\pm0.019) \times 10^{-2}$ | $1.5 \times 10^4$ | $4.0 \times 10^5$ | 1 |
| D150316-4 | Silica sand (73 μm) | No | Sieving | 58 | 33 | 57.7 | $(4.22\pm0.36) \times 10^{21}$ | $(2.0113\pm0.0045) \times 10^{-2}$ | $1.5 \times 10^4$ | $4.0 \times 10^5$ | 1 |
| D160624-1 | Silica beads (1.7 μm) | Yes [a] | Sieving | 19.9 | 17.3 | 19.6 | | | | | |
| D160624-2 | Silica beads (1.7 μm) | Yes [a] | Sieving | 19.9 | 17.3 | 19.6 | $(8.35\pm0.23) \times 10^{6}$ [b] | $(2.201\pm0.013) \times 10^{-1}$ [b] | $3.2 \times 10^3$ | $4.0 \times 10^5$ | 1 |
| D160630-1 | Silica beads (1.7 μm) | Yes [a] | Sieving | 19.9 | 17.3 | 19.6 | | | | | |
| D170429-1 | Fly ash (4.8 μm) | Yes | Sieving | 58 | 33 | 57 | $(1.649\pm0.025) \times 10^{8}$ | $(8.134\pm0.014) \times 10^{-2}$ | $1.0 \times 10^3$ | $3.9 \times 10^5$ | 2 |
| D150310-2 | Fly ash (4.8 μm) | No | Sieving | 58 | 33 | 57.7 | $(2.578\pm0.048) \times 10^{9}$ | $(6.423\pm0.013) \times 10^{-2}$ | $1.5 \times 10^4$ | $3.6 \times 10^6$ | 3 |
| D170430-3 | Glass beads (18 μm) | Yes | Sieving | 58 | 33 | 57 | $(2.951\pm0.093) \times 10^{16}$ | $(2.3496\pm0.0027) \times 10^{-2}$ | $1.0 \times 10^3$ | $3.9 \times 10^5$ | 2 |
| D150726-1 | Glass beads (18 μm) | No | Sieving | 58 | 33 | 57.7 | $(9.64\pm0.93) \times 10^{11}$ | $(4.009\pm0.023) \times 10^{-2}$ | $2.5 \times 10^3$ | $3.5 \times 10^5$ | … |
| D150728-1 | Glass beads (18 μm) | No | Sieving | 58 | 33 | 57.7 | $(9.55\pm0.79) \times 10^{11}$ | $(3.901\pm0.018) \times 10^{-2}$ | $1.5 \times 10^3$ | $2.4 \times 10^5$ | … |
| D150726-2 | Glass beads (18 μm) | No | Sieving | 58 | 33 | 57.7 | $(2.02\pm0.41) \times 10^{17}$ | $(2.222\pm0.017) \times 10^{-2}$ | $1.5 \times 10^4$ | $3.6 \times 10^6$ | … |
| D150728-2 | Glass beads (18 μm) | No | Sieving | 58 | 33 | 57.7 | $(7.4\pm1.1) \times 10^{15}$ | $(2.441\pm0.015) \times 10^{-2}$ | $1.6 \times 10^4$ | $3.6 \times 10^6$ | … |
| D190225-1 | Serpentinite (10 μm) | No | Sieving | 58.8 | 33 | 58 | $(7.42\pm0.18) \times 10^{8}$ | $(8.975\pm0.020) \times 10^{-2}$ | $2.1 \times 10^2$ | $3.6 \times 10^6$ | … |
| D180705-7 | Alumina (1.0 μm) | No | Pouring | 58 | 33 | 57 | $(6.18\pm0.28) \times 10^{8}$ | $(1.2949\pm0.0066) \times 10^{-1}$ | $3.0 \times 10^3$ | $7.8 \times 10^6$ | … |
| D151108-8 | Alumina (1.0 μm) | No | Sieving | 58 | 33 | 57.7 | $(4.04\pm0.10) \times 10^{9}$ | $(1.0930\pm0.0029) \times 10^{-1}$ | $1.5 \times 10^4$ | $3.6 \times 10^6$ | … |
| D150925-8 | Alumina (1.7 μm) | No | Sieving | 58 | 33 | 57.7 | $(1.981\pm0.016) \times 10^{8}$ | $(1.4250\pm0.0018) \times 10^{-1}$ | $1.5 \times 10^4$ | $3.6 \times 10^6$ | … |
| D151224-1 | Alumina (1.8 μm) | No | Sieving | 58 | 33 | 57.7 | $(2.15\pm0.10) \times 10^{11}$ | $(6.527\pm0.023) \times 10^{-2}$ | $1.5 \times 10^4$ | $3.6 \times 10^6$ | … |
| D151120-1 | Alumina (4.5 μm) | No | Sieving | 58 | 33 | 57.7 | $(1.83\pm0.12) \times 10^{11}$ | $(7.414\pm0.033) \times 10^{-2}$ | $1.5 \times 10^3$ | $3.6 \times 10^5$ | … |
| D150310-3 | Alumina (4.5 μm) | No | Sieving | 58 | 33 | 57.7 | $(5.53\pm0.24) \times 10^{13}$ | $(5.294\pm0.012) \times 10^{-2}$ | $1.5 \times 10^{4}$ | $3.6 \times 10^{6}$ | … |
| D150316-1 | Alumina (4.5 μm) | No | Sieving | 58 | 33 | 57.7 | $(3.39\pm0.15) \times 10^{13}$ | $(5.430\pm0.014) \times 10^{-2}$ | $1.5 \times 10^{4}$ | $3.6 \times 10^{6}$ | … |
| D170427-1 | Alumina (6.5 μm) | Yes | Sieving | 58 | 33 | 57 | $(7.36\pm0.54) \times 10^{11}$ | $(5.463\pm0.023) \times 10^{-2}$ | $1.0 \times 10^{3}$ | $3.9 \times 10^{5}$ | 2 |
| D151224-2 | Alumina (6.5 μm) | No | Sieving | 58 | 33 | 57.7 | $(1.208\pm0.083) \times 10^{16}$ | $(3.4334\pm0.0098) \times 10^{-2}$ | $1.5 \times 10^{4}$ | $3.6 \times 10^{6}$ | … |
| D170428-1 | Alumina (15 μm) | Yes | Sieving | 58 | 33 | 57 | $(3.13\pm0.22) \times 10^{12}$ | $(4.761\pm0.018) \times 10^{-2}$ | $1.0 \times 10^{3}$ | $3.9 \times 10^{5}$ | 2 |
| D151106-8 | Alumina (15 μm) | No | Sieving | 58 | 33 | 57.7 | $(6.2\pm1.1) \times 10^{18}$ | $(2.600\pm0.015) \times 10^{-2}$ | $1.5 \times 10^{4}$ | $3.6 \times 10^{6}$ | … |
| D151109-1 | Alumina (59 μm) | No | Sieving | 58 | 33 | 57.7 | $(4.1\pm1.1) \times 10^{26}$ | $(1.4058\pm0.0079) \times 10^{-2}$ | $1.6 \times 10^{4}$ | $3.6 \times 10^{6}$ | … |
| D150310-5 | Alumina (77 μm) | No | Sieving | 58 | 33 | 57.7 | $(1.49\pm0.82) \times 10^{29}$ | $(1.310\pm0.014) \times 10^{-2}$ | $3.1 \times 10^{4}$ | $3.1 \times 10^{6}$ | … |
| D150316-3 | Alumina (77 μm) | No | Sieving | 58 | 33 | 57.7 | $(1.5\pm1.1) \times 10^{35}$ | $(1.010\pm0.011) \times 10^{-2}$ | $1.6 \times 10^{4}$ | $2.4 \times 10^{6}$ | 3 |

[a] We used a new sample powder opened within 1 week after purchase.

[b] Values are obtained from an average compression curve of three measurements.

**References.** (1) Omura and Nakamura (2018), (2) Omura and Nakamura (2017), (3) Omura et al. (2016).

**Table 3 Information and Results of Mixture Samples**

| | | | | Sample container | | Piston | | | | |
|---|---|---|---|---|---|---|---|---|---|---|
| No. | Dust | Dried Particles Yes/No | Dust Volume Fraction | Inner Diameter (Mm) | Depth (Mm) | Diameter (Mm) | Initial Porosity | Final Porosity[a] | *K'* | *n* |
| 181113A-1 | Fly ash (4.8 μm) | No | 1 | 58.8 | 33 | 58 | 0.60 | 0.40 | $(2.016 \pm 0.078) \times 10^{10}$ | $(6.069 \pm 0.018) \times 10^{-2}$ |
| 181119A-2 | Fly ash (4.8 μm) | No | 0.72 | 58.8 | 33 | 58 | 0.58 | 0.33 | $(1.514 \pm 0.090) \times 10^{9}$ | $(6.156 \pm 0.032) \times 10^{-2}$ |
| 190201A-1 | Fly ash (4.8 μm) | No | 0.72 | 58.8 | 33 | 58 | 0.58 | 0.33 | $(2.27 \pm 0.11) \times 10^{9}$ | $(5.988 \pm 0.024) \times 10^{-2}$ |
| 190204A-1 | Fly ash (4.8 μm) | No | 0.72 | 58.8 | 33 | 58 | 0.59 | 0.34 | $(2.03 \pm 0.10) \times 10^{9}$ | $(6.098 \pm 0.026) \times 10^{-2}$ |
| 181119A-3 | Fly ash (4.8 μm) | No | 0.52 | 58.8 | 33 | 58 | 0.51 | 0.27 | $(1.018 \pm 0.055) \times 10^{9}$ | $(5.128 \pm 0.024) \times 10^{-2}$ |
| 190201A-2 | Fly ash (4.8 μm) | No | 0.52 | 58.8 | 33 | 58 | 0.52 | 0.28 | $(1.376 \pm 0.066) \times 10^{9}$ | $(5.100 \pm 0.021) \times 10^{-2}$ |
| 190204A-2 | Fly ash (4.8 μm) | No | 0.52 | 58.8 | 33 | 58 | 0.51 | 0.28 | $(1.340 \pm 0.073) \times 10^{9}$ | $(5.037 \pm 0.024) \times 10^{-2}$ |
| 181119A-4 | Fly ash (4.8 μm) | No | 0.37 | 58.8 | 33 | 58 | 0.46 | 0.23 | $(7.32 \pm 0.49) \times 10^{8}$ | $(4.307 \pm 0.026) \times 10^{-2}$ |
| 190201A-3 | Fly ash (4.8 μm) | No | 0.37 | 58.8 | 33 | 58 | 0.44 | 0.22 | $(1.367 \pm 0.079) \times 10^{9}$ | $(3.925 \pm 0.020) \times 10^{-2}$ |
| 190204A-3 | Fly ash (4.8 μm) | No | 0.37 | 58.8 | 33 | 58 | 0.45 | 0.24 | $(9.37 \pm 0.65) \times 10^{8}$ | $(4.255 \pm 0.026) \times 10^{-2}$ |
| 190201A-4 | Fly ash (4.8 μm) | No | 0.21 | 58.8 | 33 | 58 | 0.39 | 0.25 | $(3.64 \pm 0.59) \times 10^{10}$ | $(2.732 \pm 0.029) \times 10^{-2}$ |
| 190204A-4 | Fly ash (4.8 μm) | No | 0.21 | 58.8 | 33 | 58 | 0.39 | 0.25 | $(9.2 \pm 1.7) \times 10^{10}$ | $(2.579 \pm 0.030) \times 10^{-2}$ |
| 190204A-5 | Fly ash (4.8 μm) | No | 0.21 | 58.8 | 33 | 58 | 0.39 | 0.25 | $(7.7 \pm 1.4) \times 10^{10}$ | $(2.605 \pm 0.030) \times 10^{-2}$ |
| 181119A-1 | Silica powder (1.5 μm) | No | 1 | 58.8 | 33 | 58 | 0.81 | 0.47 | $(1.7399 \pm 0.0070) \times 10^{9}$ | $(1.13539 \pm 0.00043) \times 10^{-1}$ |
| 181117A-2 | Silica powder (1.5 μm) | No | 0.73 | 58.8 | 33 | 55.4 | 0.77 | 0.40 | $(8.239 \pm 0.075) \times 10^{8}$ | $(1.03221 \pm 0.00092) \times 10^{-1}$ |
| 190301A-1 | Silica powder (1.5 μm) | No | 0.73 | 58.8 | 33 | 55.4 | 0.75 | 0.41 | $(1.485 \pm 0.011) \times 10^{9}$ | $(9.7202 \pm 0.0069) \times 10^{-2}$ |
| 190301A-4 | Silica powder (1.5 μm) | No | 0.73 | 58.8 | 33 | 55.4 | 0.75 | 0.41 | $(1.541 \pm 0.014) \times 10^{9}$ | $(9.6054 \pm 0.0080) \times 10^{-2}$ |
| 181117A-1 | Silica powder (1.5 μm) | No | 0.53 | 58.8 | 33 | 55.4 | 0.71 | 0.36 | $(6.273 \pm 0.097) \times 10^{8}$ | $(8.957 \pm 0.014) \times 10^{-2}$ |
| 190301A-2 | Silica powder (1.5 μm) | No | 0.53 | 58.8 | 33 | 55.4 | 0.71 | 0.36 | $(8.36 \pm 0.10) \times 10^{8}$ | $(8.825 \pm 0.010) \times 10^{-2}$ |
| 190301A-5 | Silica powder (1.5 μm) | No | 0.53 | 58.8 | 33 | 55.4 | 0.71 | 0.36 | $(8.94 \pm 0.11) \times 10^{8}$ | $(8.622 \pm 0.010) \times 10^{-2}$ |
| 181117A-3 | Silica powder (1.5 μm) | No | 0.38 | 58.8 | 33 | 55.4 | 0.65 | 0.33 | $(8.00 \pm 0.21) \times 10^{8}$ | $(7.311 \pm 0.018) \times 10^{-2}$ |
| 190301A-3 | Silica powder (1.5 μm) | No | 0.38 | 58.8 | 33 | 55.4 | 0.66 | 0.33 | $(8.95 \pm 0.19) \times 10^{8}$ | $(7.439 \pm 0.015) \times 10^{-2}$ |
| 190301A-6 | Silica powder (1.5 μm) | No | 0.38 | 58.8 | 33 | 55.4 | 0.65 | 0.34 | $(1.174 \pm 0.025) \times 10^{9}$ | $(7.193 \pm 0.014) \times 10^{-2}$ |
| 181127A-2 | Silica powder (1.5 μm) | No | 0.22 | 58.8 | 33 | 58 | 0.58 | 0.33 | $(3.16 \pm 0.20) \times 10^{9}$ | $(5.341 \pm 0.027) \times 10^{-2}$ |
| 190301A-7 | Silica powder (1.5 μm) | No | 0.22 | 58.8 | 33 | 58 | 0.55 | 0.34 | $(3.90 \pm 0.24) \times 10^{10}$ | $(4.265 \pm 0.019) \times 10^{-2}$ |
| 190301A-8 | Silica powder (1.5 μm) | No | 0.22 | 58.8 | 33 | 58 | 0.56 | 0.33 | $(1.81 \pm 0.11) \times 10^{10}$ | $(4.551 \pm 0.021) \times 10^{-2}$ |

**Note.**

[a] The porosity of the sample under the pressure of $5 \times 10^{6}$ Pa.

## 3. INTERNAL POROSITY STRUCTURE OF CHONDRITE PARENT BODIES

We calculated the internal porosity structure of primordial porous chondrite parent bodies with a given dust volume fraction. The internal porosity structure of a primordial chondrite parent body, which is continuous and in hydrostatic equilibrium, can be calculated using the polytropic relationship of the constituent material and the Lane–Emden equation (Omura & Nakamura 2018). The polytropic relationship used in the Lane–Emden equation (e.g., Chandrasekhar 1957) is

$$P = K\rho^{(n+1)/n}, \tag{9}$$

where $P$ is the pressure, $K$ is a constant, $\rho$ is the density, and $n$ is the polytropic index. To introduce a constant $K'$ with a pressure dimension, Equation (9) can be modified as

$$P = K'\left(\frac{\rho}{\rho_\mathrm{a}}\right)^{(n+1)/n} = K' f^{(n+1)/n}. \tag{10}$$

To solve the Lane–Emden equation, $K$ was calculated given $K'$, $\rho_\mathrm{a}$, and $n$. Discussions about values of $K'$ and $n$ for each sample are provided in the Appendix. The values of $n$ and $K'$ used for the calculation were obtained by interpolating or averaging the results shown in Figures A1c and d and are listed in Table 4.

**Table 4 Values Used for Calculation**

| Dust | Dust Volume Fraction | $n$ | $K'$ | $\rho$ (kgm$^{-3}$) | $K$ | Remarks |
|---|---|---|---|---|---|---|
| Silica powder (1.5μm) | 1.00 | $1.135 \times 10^{-1}$ | $1.7 \times 10^{9}$ | 2200 | $2.9 \times 10^{-24}$ | 1 |
| Silica powder (1.5μm) | 0.70 | $9.713 \times 10^{-2}$ | $1.2 \times 10^{9}$ | 2290 | $1.3 \times 10^{-29}$ | 2 |
| Silica powder (1.5μm) | 0.40 | $7.482 \times 10^{-2}$ | $9.2 \times 10^{8}$ | 2380 | $2.9 \times 10^{-40}$ | 2 |
| Silica powder (1.5μm) | 0.22 | $4.677 \times 10^{-2}$ | $1.3 \times 10^{10}$ | 2434 | $2.1 \times 10^{-66}$ | 3 |
| Fly ash (4.8 μm) | 1.00 | $6.069 \times 10^{-2}$ | $2.0 \times 10^{10}$ | 2300 | $3.6 \times 10^{-49}$ | 1 |
| Fly ash (4.8 μm) | 0.70 | $5.978 \times 10^{-2}$ | $1.8 \times 10^{9}$ | 2360 | $2.9 \times 10^{-51}$ | 2 |
| Fly ash (4.8 μm) | 0.40 | $4.316 \times 10^{-2}$ | $1.0 \times 10^{9}$ | 2420 | $1.7 \times 10^{-73}$ | 2 |
| Fly ash (4.8 μm) | 0.21 | $2.637 \times 10^{-2}$ | $6.4 \times 10^{10}$ | 2457 | $6.8 \times 10^{-122}$ | 3 |

**Notes.** (1) The values are obtained by the result of an experiment. (2). The values are obtained by interpolating the results shown in Table 3nd Figures A1(c) and (d). (3) The values are an average of the results of three experiments.

Figure 3 shows the calculated internal porosity structures of primordial chondrite parent bodies with radii of 10, 30, and 100 km. Porosity can be calculated from density as

$$\varepsilon = 1 - \frac{\rho}{\rho_{\mathrm{a}}}. \tag{11}$$

Note that the porosity structure estimated in this study is that of primordial chondrite parent bodies formed by accretion of constituent particles of chondrites. The porosity decreases with increasing distance from the surface and radius of the body. The relationship between the radius and the calculated bulk porosity of chondrite parent bodies obtained from the results for 1.5 μm dust samples is shown in Figure 4. For the sake of our calculation, we set the porosity to 1 at the surface. Of course the porosity immediately below the surface will evolve with impacts and thermal and aqueous alteration.

Bodies with a lower dust volume fraction have lower porosities. This tendency is consistent with the porosity of carbonaceous chondrites being higher than that of ordinary chondrites, although their thermal history should also have affected the porosity. Bodies consisting of the smaller-sized dust particles have higher porosity in Figure 3. The matrix particles may be smaller than the diameters of the dust particles used in this study (1.5 μm). The measured average size of matrix particles of the CV chondrites of Allende, Kava, and Vigarano is 0.96, 0.5, and 0.5 μm, respectively (Forman et al. 2019). Moreover, if the primordial body consists of finer dust particles similar to CP IDPs, the porosity structure of the body could be different from that estimated in this study. Estimation of the compaction behavior of such particles is beyond the scope of this study, but it is an important future subject. The expected increase in the interparticle force in the vacuum environment (Steinpilz et al. 2019) may increase porosity.

Figure 3 shows the porosities of chondrites (with matrix volume fractions corresponding to each dust volume fraction). Although the porosities of the parent bodies of CI and CM chondrites, the chondrites that underwent aqueous alteration, are likely to have been affected by the mixing of ice particles, this is not considered in current estimations. However, the calculated porosity obtained from the results of the 1.5 μm dust sample is higher than the chondrite porosity but consistent with the estimated porosity of a boulder on Ryugu (Grott et al. 2019). The macroporosity of asteroids estimated using the density of chondrites may be overestimated. The difference between the calculated porosity and the porosity of the meteorites may have been caused by thermal processes, such as hot pressing (Henke et al. 2012) and impact-induced compaction and heating ( Hirata et al. 2009; Beitz et al. 2013). For instance, the deformation microstructures detected in Allende suggest impact-induced compaction of a highly porous initial material (Forman et al. 2016). Additionally, aqueous alteration may have had an effect on porosity.

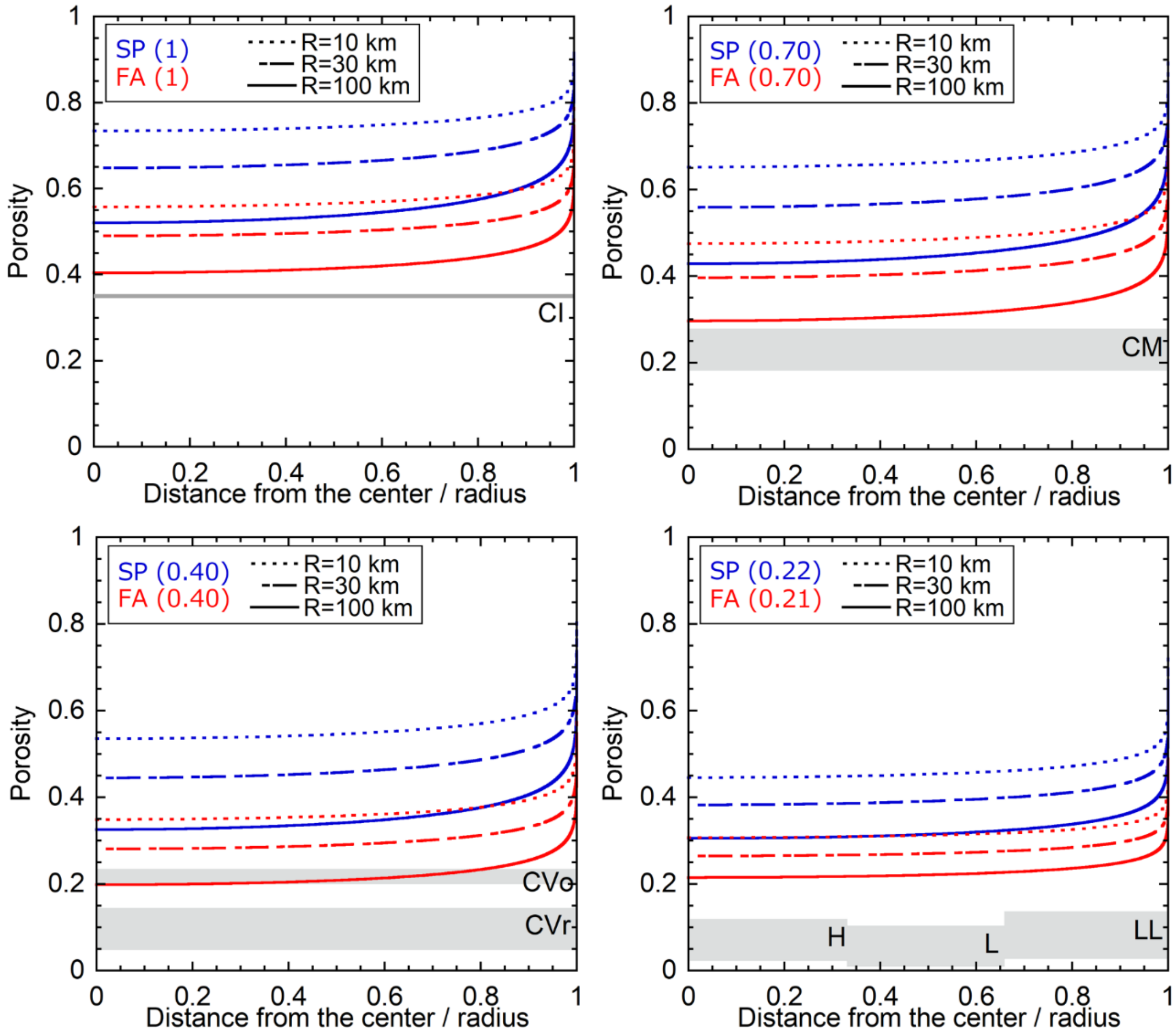

**Figure 3.** Calculation results of internal porosity structure of the primordial chondrite parent body. The dust volume fraction is shown at the top left of the graph. The blue and red curves correspond to the results with silica powder (1.5 μm) and fly ash (4.8 μm), respectively. Dotted, dashed, and solid curves show the calculation results of the bodies with radii of 10, 30, and 100 km, respectively. The porosity ranges of chondrites, with the volume fraction of the matrix corresponding to each dust volume fraction, are shown using gray bands (where the lengths and horizontal positions of the bands are arbitrary). The chondrule sizes of the chondrites shown in this figure range from 0.3 to 1.0, although the chondrule size of CI-group chondrites is not known (Weisberg et al. 2006).

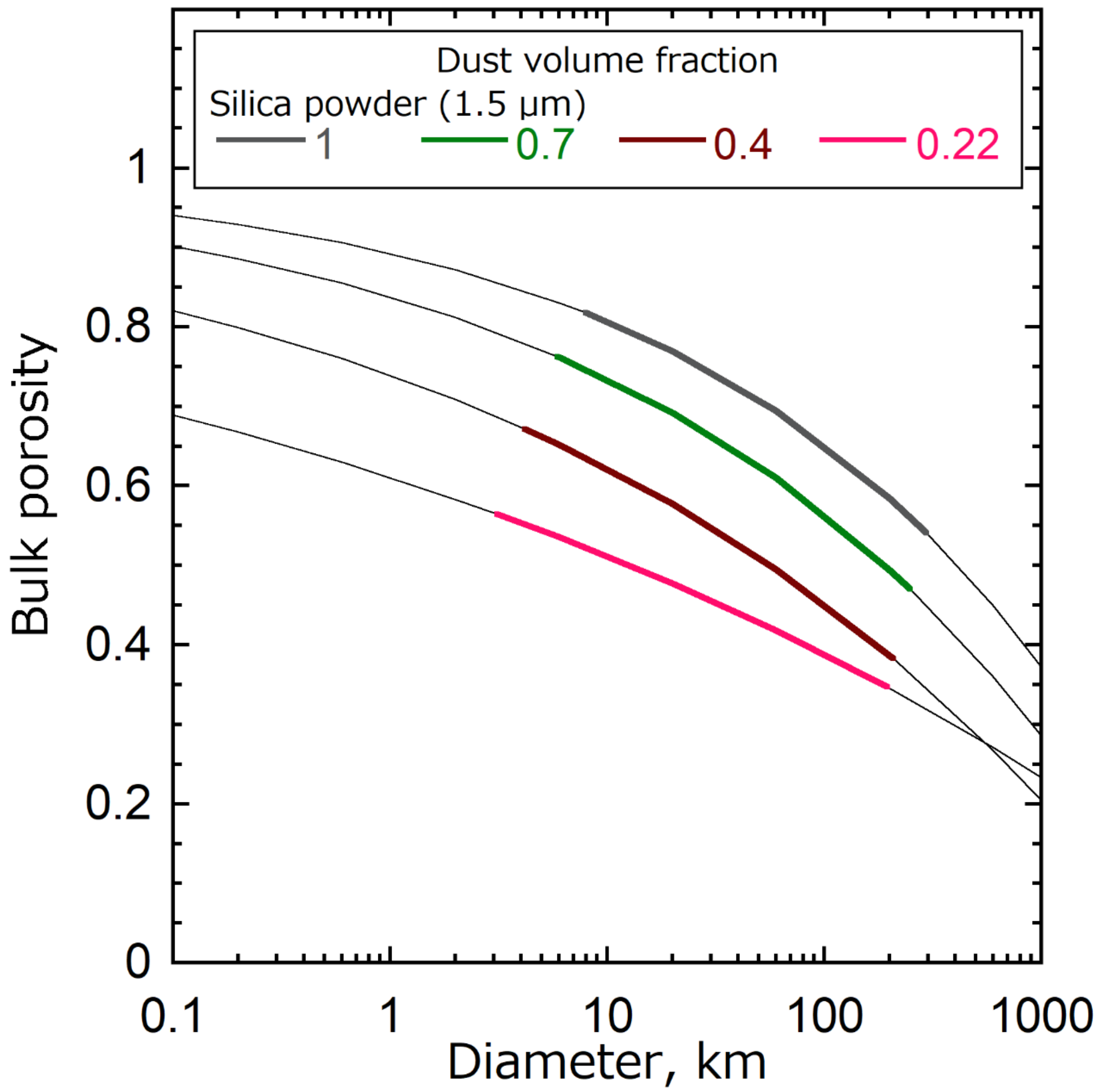

**Figure 4**. Calculated bulk porosity of primordial chondrite parent bodies. The curves show the calculation results, and the color of the curve corresponds to the dust volume fraction. The thin black curves indicate that the pressure at the center of the body is outside the range of parameter fitting in this study.

## 4. Summary

We conducted compaction experiments using chondrite component analogs (i.e., dust and dust–beads mixture samples) and obtained a relationship between the applied pressure and the filling factor for each sample. Further, we measured the effect of the dust volume fraction on the compaction curve for mixture samples and discussed the result based on a modified two-layer model. Our results showed that the compaction behavior of a sample is dominated by that of the dust when the dust volume fraction is larger than ~0.5. The compaction curve of each sample was approximated with a power-law form: $P = K'f^{(n+1)/n}$. The values of $K'$ and $n$ for mixture samples were used to calculate the internal porosity structure of chondrite parent bodies.

The calculated porosity was higher than the chondrite porosity. The estimated porosity range matches that of the Ryugu-boulder's porosity, which may suggest that the boulder represents the microscopic structure of a primordial planetesimal. Note that the dust particle size of our sample was probably larger than that of primitive matrix particles. Moreover, porosity tends to decrease as the dust volume fraction decreases. The relatively higher volume fraction of the matrix may be one of the reasons why most meteorites with high porosity are carbonaceous chondrites. The most porous structure of the primitive bodies provided in this study can be employed as a basis for future studies on the evolution of small bodies and the solar system. However, the matrix analogs used in this study are not a perfect analog, and this study is a first step to estimating the most porous structure. Future studies will consider bodies that consist of more primitive (CP IDP-like) dust with perhaps hierarchical structure.

We thank M. Hyodo for providing access to the laser diffractometer and M. Suzuki for their advice on handling of the powder material. We also thank H. Yoneda for helping us with the writing of the calculation cord. A preliminary compression experiment was also performed using a compressive testing machine installed at the Hypervelocity Impact Facility (former facility name: The Space Plasma Laboratory), ISAS, JAXA, Japan. We thank S. Hasegawa for providing access to the compressive testing machine. Finally, we are grateful to anonymous reviewers for constructive comments on this paper. This research was supported by the Hosokawa Powder Technology Foundation and KAKENHI from the Japan Society for the Promotion of Science (JSPS; grant No. 18K03723).

## APPENDIX: FITTING OF COMPACTION CURVES

We approximated the compaction curves of the dust-only and mixture samples using Equation (10). The values of $n$ and $K'$ obtained for each curve are shown in Tables 2 and 3. For the dust-only samples, a pressure range in which the data could be well approximated by a power law was used for fitting. The minimum and maximum pressures of the fitting range are shown in Table 2, for which hydrostatic pressure due to the powders can be neglected. For the mixture samples, the results at pressures higher than the yield strength were used for fitting.

The relationships between the $n$ or $K'$ values of the dust samples and the median diameter (Figure 1, Table 1) of the sample powder are shown in Figures A1(a) and (b). The values obtained by the approximation of the compaction curves in our previous studies (Omura et al. 2016; Omura & Nakamura 2017, 2018) are also included. Both $n$ and $K'$ of silica and silicate powders primarily vary depending on the median diameter of the powders, irrespective of the composition and particle shape, whereas alumina powders tend to have smaller $n$ and larger $K'$. No significant correlation is shown between the broadness of the particle size distribution and the values of $n$ or $K'$. The systematic decrease and increase of $n$ and $K'$, respectively, along the median particle size show that the compaction behavior of the chondrite matrix can be discussed based on median particle size. The relationships between the values of $(n+1)/n$ and log $K'$ in the mixture samples (Figure 2) and the volume fraction of dust are shown in Figures A1(c) and (d). Each value is normalized by that of the sample with a dust volume fraction of 1 (dust-only samples). The normalized value of $(n+1)/n$ ranges from 1 to ~2.5, irrespective of the dust material, and decreases with the dust volume fraction. The normalized log $K'$ tends to gradually increase with the dust volume fraction for the dust-rich sample (dust volume fraction >~0.4), whereas the normalized log $K'$ for the dust-poor sample (dust volume fraction of ~0.2) is much larger than that extrapolated from those of the dust-rich samples, suggesting that the compaction behavior of the dust-poor sample is different from that of the dust-rich sample, as indicated in Figure 2.

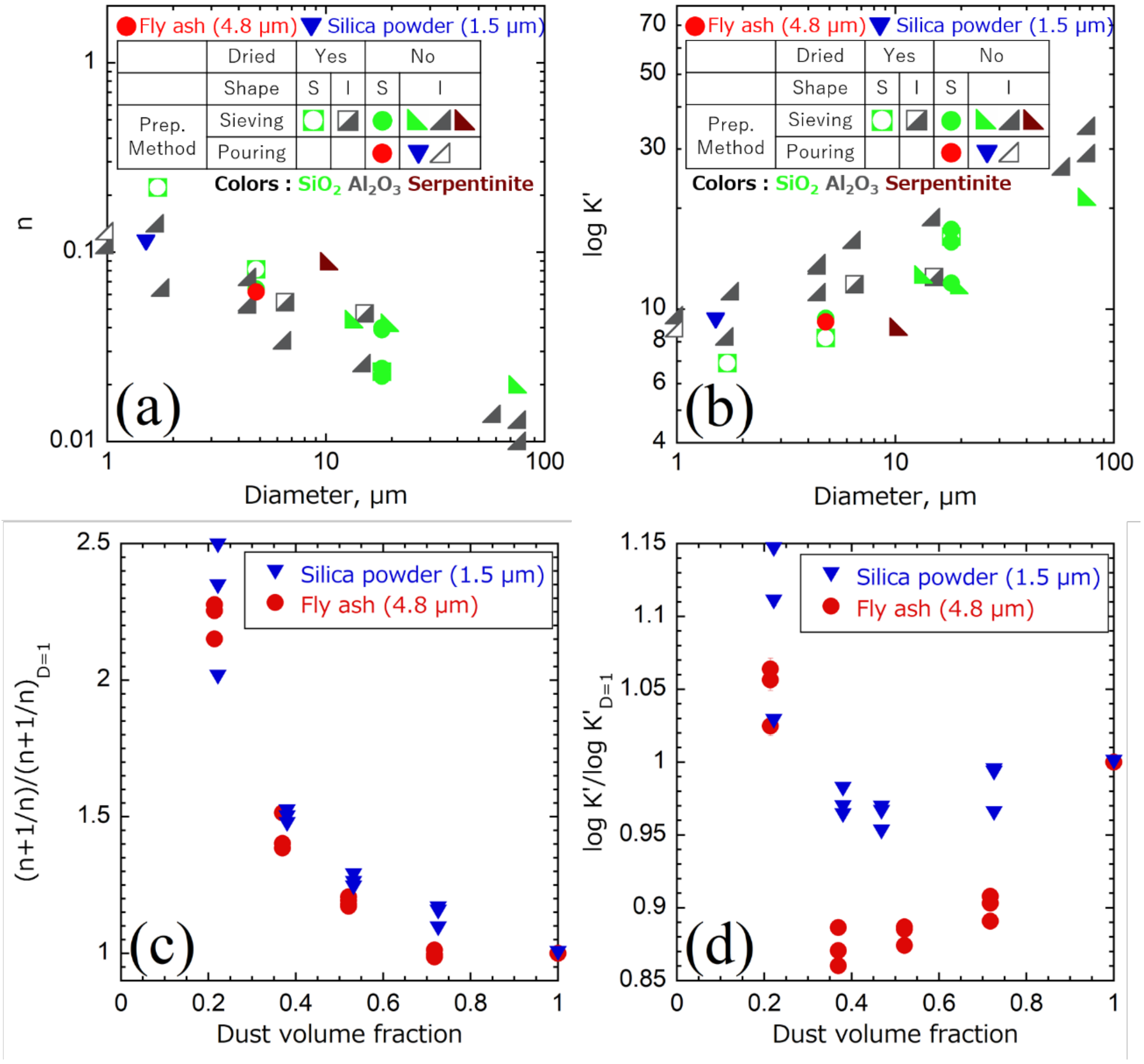

**Figure A1.** Polytropic index $n$ and constant $K'$ obtained by the approximation of compaction curves. (a) and (b) Particle size dependence of (a) the polytropic index, $n$, and (b) the constant $K'$ for dust samples. Each symbol corresponds to the sample condition (dried or not), particle shape (S: spherical; I: irregular), and preparation method used (sieved or poured). The symbol colors correspond to the major material components of each sample. The dust materials used in the mixture samples are shown by blue triangles and red spheres. (c) and (d) Dust volume fraction dependence of (c) $(n+1)/n$ and (d) $\log K'$ for the mixture samples. The triangle and sphere correspond to 1.5 μm and 4.8 μm dust particles, respectively.